\documentclass[prb,twocolumn,showpacs,superscriptaddress,unsortedaddress]{revtex4}
\usepackage{amsfonts}
\usepackage{amsmath}
\usepackage{amssymb}
\usepackage{graphicx}
\usepackage{dcolumn}
\usepackage{bm}% bold math
\usepackage{color}

\begin{document}

\title{Muon spin
rotation study of the magnetic penetration depth in the
intercalated graphite superconductor CaC$_6$}
\author{D. Di Castro}
\affiliation{CNR-SPIN and Dipartimento di Ingegneria Meccanica, Universit\`a di Roma Tor Vergata, Via del
Politecnico 1, I-00133 Roma, Italy}
%\affiliation{Physik-Institut der
%Universit\"{a}t Z\"{u}rich, Winterthurerstrasse 190, CH-8057
%Z\"urich, Switzerland}
\author{A. Kanigel} \affiliation{Department of
Physics, Technion, Haifa 32000, Israel}
\affiliation{Department of
Physics, University of Illinois at Chicago, Chicago, IL 60607}
\author{A. Maisuradze}
\affiliation{Laboratory for Muon Spin Spectroscopy, Paul Scherrer
Institut, CH-5232 Villigen PSI, Switzerland}
\author{A. Keren}
\affiliation{Department of Physics, Technion, Haifa 32000, Israel}
\author{P. Postorino}
\affiliation{Dipartimento di Fisica, Universit\`a di Roma La
Sapienza, Piazzale Aldo Moro 2, I-00185 Roma, Italy}
\author{D. Rosenmann}
\affiliation{Center for Nanoscale Materials, Argonne National Laboratory, Argonne, IL 60439}
\author{U. Welp}
\affiliation{Materials Science Division, Argonne National Laboratory, Argonne, IL 60439}
\author{G. Karapetrov}
\affiliation{Materials Science Division, Argonne National Laboratory, Argonne, IL 60439}
\author{H. Claus}
\affiliation{Materials Science Division, Argonne National Laboratory, Argonne, IL 60439}
\author{D.G. Hinks}
\affiliation{Materials Science Division, Argonne National Laboratory, Argonne, IL 60439}
\author{A. Amato}
\affiliation{Laboratory for Muon Spin Spectroscopy, Paul Scherrer
Institut, CH-5232 Villigen PSI, Switzerland}
\author{J. C. Campuzano}
\affiliation{Department of Physics, University of Illinois at Chicago, Chicago, IL 60607}
\affiliation{Materials Science Division, Argonne National Laboratory, Argonne, IL 60439}

\begin{abstract}
We report temperature- and magnetic field-dependent bulk muon spin
rotation measurements in a c-axis oriented superconductor
CaC$_{6}$ in the mixed state. Using both a simple second moment
analysis and the more precise analytical Ginzburg-Landau model, we
obtained a field independent in-plane magnetic penetration depth
$\lambda_{ab}(0)$ = 72(3) nm. The temperature dependencies of the
normalized muon spin relaxation rate and of the normalized
superfluid density result to be identical, and both are well
represented by the clean limit BCS model with 2$\Delta$/k$_BT_c$ =
3.6(1), suggesting that CaC$_6$ is a fully gapped BCS
superconductor in the clean limit regime.
\end{abstract}

\date{\today }
\maketitle

Recently the field of graphite intercalated compounds (GICs) has regained
attention after the discovery of the  superconducting GIC CaC$_{6}$ with T%
$_{c}\sim $11.5K.\cite{Weller,Emery} Soon after the discovery it
was suggested \cite{csanyi} that CaC$_{6}$ is an unconventional
superconductor, and that the pairing interaction is of electronic
origin. As time elapsed, both calculations \cite{Theory} and
experiments showed that CaC$_{6}$ is a conventional BCS
superconductor. In particular, experiments revealed the existence
of a significant Ca isotope effect,\cite{hinks_iso} the absence of
gap nodes,\cite{CP} a single gap \cite{Bergeal} with BCS
temperature dependence,\cite{Gonnelli} as well as the BCS
temperature dependence of the London penetration depth
.\cite{Lamura} Moreover, calculations,\cite{Theory} supported by
experimental evidence
,\cite{CP,Bergeal,ANL_mag,Sanna,Sugawara,Okazaki} showed the
importance of a Ca derived interlayer band, crossing the Fermi
energy, which have sufficiently strong coupling with both in-plane
intercalant and out-of-plane graphite phonon modes to explain the
T$_c$ of CaC$_6$ within a standard electron-phonon coupling
mechanism. On the other hand, the coupling between
graphene-derived electrons and high-frequency graphene-derived
phonons was also  demonstrated to be  relevant.\cite{Valla}

At present the mechanism of superconductivity in CaC$_6$ and most
of its properties have been investigated and clarified.
Nevertheless a complete  description of the superconducting state
in a Type II superconductor requires a whole coherent set  of
parameters measured with high precision, such as the value of the
coherence length $\xi $  and the magnetic penetration depth
$\lambda$. These points are somehow still lacking. Indeed, two
very different values of the in plane penetration depth lambda
$\lambda_{ab}(0)$ have been reported in the literature
\cite{Lamura,Cubitt} and the temperature dependent  superfluid
density, $\rho(T)$, has been measured with surface sensitive
technique only.\cite{Lamura}  For the extremely air sensitive
superconductor CaC$_6$,  bulk measurement of the superconducting
properties would be thus very important. We also notice that a new
unambiguous bulk determination of $\rho(T)$ could also help in
clarifying the nature of the CaC$_6$ conduction regime, since both
dirty limit \cite{Lamura,Nagel,Muranyi} and clean limit
\cite{Mialitsin,Cubitt} regime have been reported so far. This
system is also interesting from the point of view of $\mu$SR data
analysis, since it is a low $\kappa = \lambda/\xi$ superconductor
with reduced, in field, vortex phase.

 The  muon spin rotation ($\mu $SR) technique is known to be one
of the most indicated techniques to measure the bulk properties of
type II superconductors in the vortex state. In this paper we
report a transverse field muon spin rotation (TF-$\mu $SR)
experiment on high quality CaC$_{6}$ c-axis oriented samples,
aimed to probe the vortex lattice and determine $\xi_{ab} (0)$,
$\lambda_{ab}(0)$, and the temperature dependence of the
superfluid density $\rho(T)$.\cite{note} By using a peculiar
method developed to collect reliable $\mu$SR data  on very thin
samples, such as the CaC$_6$ ones, we obtained $\lambda_{ab}(0)$ =
72(3) nm and  $\xi_{ab}(0) \simeq$ 38 nm. The  penetration depth
results to be field independent, suggesting a fully gapped Fermi
surface. The
 temperature dependence of the normalized superfluid
density is very similar to that one of the normalized muon spin
relaxation rate, as it is expected for a $\kappa \simeq 2$
superconductor at fields  about $B_{c2}/2$. Both the temperature
dependencies  are well represented by the clean limit BCS model,
with 2$\Delta$/k$_BT_c$ = 3.6(1), suggesting that CaC$_6$ is a BCS
superconductor in the clean limit regime.

In  TF-$\mu $SR the sample is cooled-down in a magnetic field,
which here was parallel to the c-axis. The muons are implanted in
the sample with their initial spin polarization almost
perpendicular to the external magnetic field. Each muon precesses
around to the local field. When the superconducting sample is in
the mixed state, each muon probes a slightly different field. As a
result, a de-phasing process takes place and the average muon spin
polarization decays. From the average polarization of the muon
as a function of time, the parameters of the vortex lattice, $\lambda $ and $%
\xi $, can be extracted.\cite{Mais,Sonier_review}

CaC$_{6}$ samples were prepared using the alloy method as
described by Emery \textit{et al.}.\cite{Emery} A stainless steel
tube is loaded with Lithium and Calcium in the ratio 3:1. Natural
graphite flakes are then added to the ampoule. The reaction takes
place in an argon atmosphere for 10 days at 350 C. The flakes are
typically of a size 2~mm $\times $ 2~mm, with a thickness of
~50$~\mu $m. The low-field superconducting transition onset is at
$T_c$ = 11.6 K.\cite{Hinks-Tc}  It was reported previously that
these samples are very reactive and tend to degrade very fast.
This is observed as an increase in the transition width and a
decrease in the superconducting volume fraction. For that reason
the samples were kept in an He atmosphere and overall were exposed
to the room atmosphere for not more than a few minutes.

\begin{figure}[tbp]
\vspace{0.5cm}
\par
\begin{center}
\includegraphics[width=7cm]{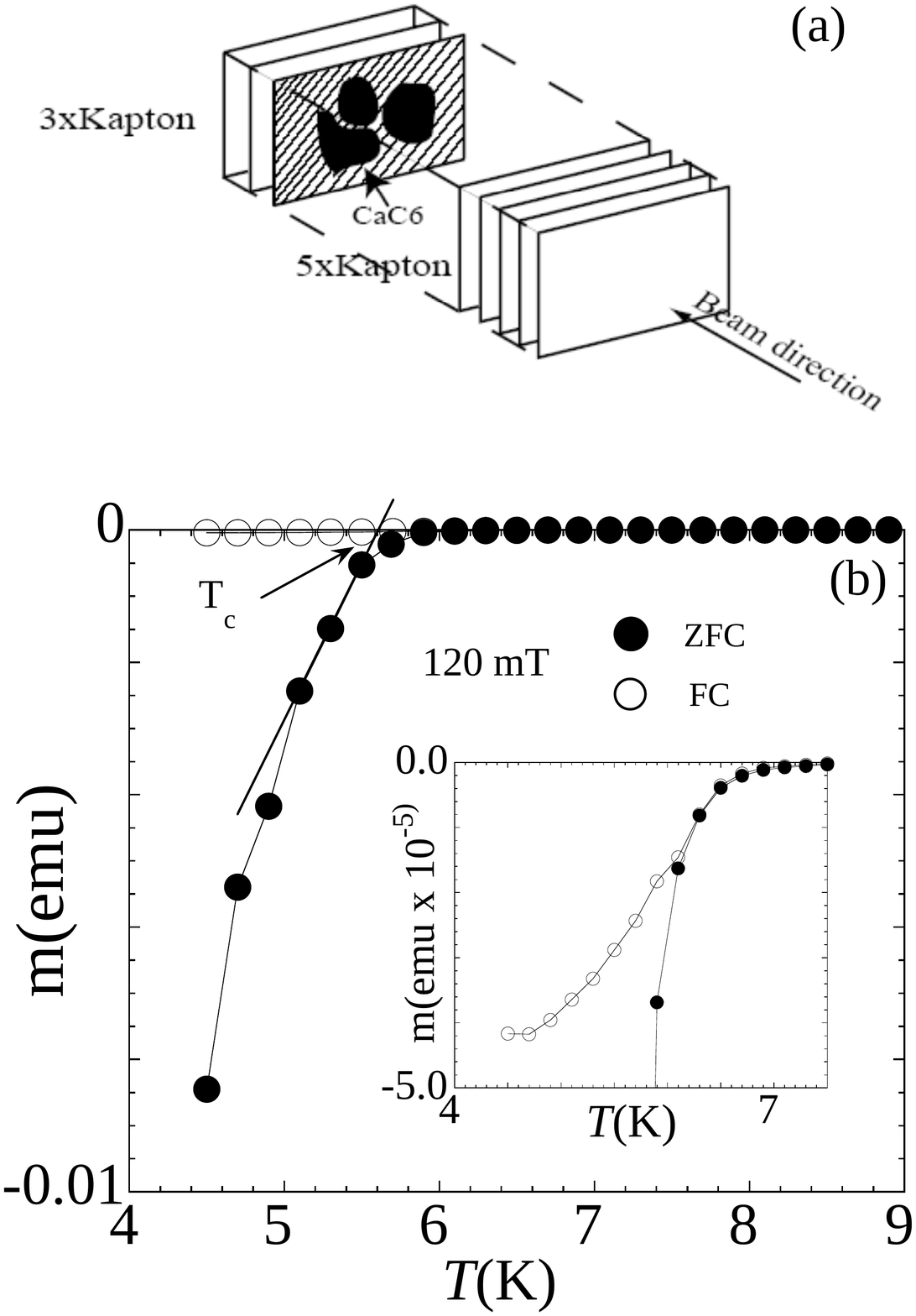}
\end{center}
\caption{(a) Illustration of the CaC$_{6}$ sample and the Kapton
stacks used to degrade the muon beam (see text). (b) Magnetic
moment of CaC$_6$   as a function of temperature measured at the
same field used in TF-$\mu $SR measurements (120 mT). The closed
(open) symbols represent zero-field cooling (field cooling)
measurements. Inset to panel (b): magnetic moment on enlarged
scale to better show the ZFC data.} \label{figure1}
\end{figure}

The experiments were done in the GPS spectrometer at the Paul
Scherrer Institute  PSI Villigen, Switzerland. Three flakes were
used, covering together an area of about 4~mm $\times $ 4~mm. The
CaC$_{6}$ samples are too thin to stop the muons. Given the muon
stopping distance and the density of CaC$_6$, a sample of a few
hundred microns is needed for this purpose.
 To maximize the number of muons stopping in the samples, we used a
method developed to measure very thin crystals.\cite{Kapton} The
samples were sandwiched between two stacks of Kapton foil
rectangles. This assembly is illustrated in the panel (a) of
Fig.~\ref{figure1}. Each Kapton foil is 125$~\mu $m thick. The
role of the Kapton foils in front is to slow down the muons so
that most of them will stop in the sample. The typical statistics
of our $\mu$SR measurements was about 17 millions events.

Muons in Kapton are known to have a very high probability of
forming muonium (muon-electron bound state) However, a muonium
signal is not observed in our experiment. In addition, muons
stopped by Kapton do not contribute substantially to the measured
asymmetry since it is known  that the asymmetry is strongly
reduced by a thick Kapton layer. \cite{Kapton} The Kapton foils in
the back of the sample prevent slow muons that did not stop in the
sample from stopping in the windows of the cryostat. In addition,
we used a veto detector, that reduces the background signal by
removing events from the data where the incoming muon missed the
sample. Using the veto counter and Kapton combination we were able
to get an asymmetry of about 0.1, about half of that we estimated
to be "good" signal coming from the CaC$_{6}$ sample.

\begin{figure}[tbp]
\vspace{0.5cm}
\par
\begin{center}
\includegraphics[width=8.5cm]{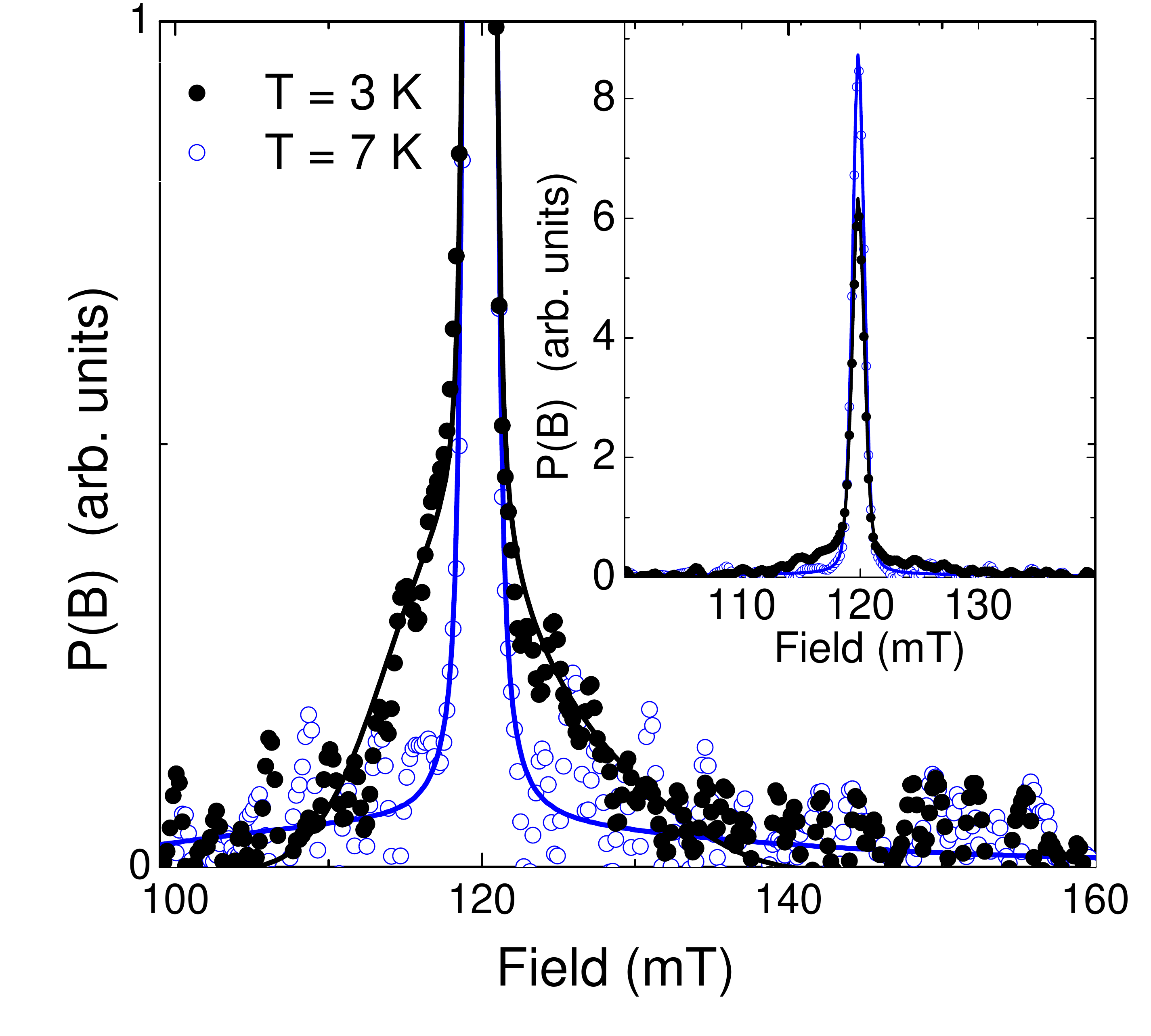}
\end{center}
\caption{Magnetic field distribution $P(B)$  at an applied field
of 120 mT above (7 K) and below (3 K) $T_c$ obtained from the
measured $\mu$SR time spectra by means of fast Fourier Transform.
The solid lines are fit by using Eq.~\ref{eq:TdomainSpec}. Inset:
the same as in the main panel on larger scales.} \label{FFT}
\end{figure}

The $\mu$SR signal was recorded in the usual time-differential way
by counting positrons from decaying muons as a function of time.
The time dependence of the positron rate is given by the
expression :\cite{Schenck}

\begin{equation}
 \label{positronRate}
\
N(t)=N_0\frac{1}{\tau_{\mu}}e^{-\frac{t}{\tau_{\mu}}}[1+aP(t)]+N_{bg}
\end{equation}

\noindent where $N_0$ is the normalization constant, $N_{bg}$
denotes the time-independent background,
$\tau_{\mu}$=2.19703(4)x10$^{-6}$ s is the muon lifetime, $a$ is
the maximum decay asymmetry for the particular detector telescope
 and $P(t)$ is the polarization of the muon
ensemble $P(t) = \int P(B)\cos(\gamma_{\mu}Bt+\phi)dB$. Here
$\gamma_{\mu} =2\pi$ x $135.5342$ MHz/T is the muon gyromagnetic
ratio and $\phi$ is the angle between the initial muon
polarization and the effective symmetry axis of a positron
detector. $P(t)$ can be linked to the internal field distribution
$P(B)$ by using the algorithm of Fourier transform.\cite{Schenck}

 In Fig.~\ref{FFT} the magnetic
field distribution $P(B)$  at an applied field of 120 mT above (7
K) and below (3 K) $T_c$, obtained from the measured $\mu$SR time
spectra by performing fast Fourier Transform, is shown. In the
normal state, $P(B)$ is a sharp symmetric line centered at the
position of the external magnetic field. Below $T_c$ the field
distribution results to be composed by two signals: one is
represented by a sharp symmetric peak centered at the position of
the external field and ascribed to different sources of background
(cryostat window, Kapton foils), and the other is a broadened
line, which signals the formation of a vortex lattice (VL) in the
superconducting part of the sample. Indeed, VL makes the field in
the sample highly inhomogeneous. This induces an increase of
 the muon spin relaxation rate $\sigma$, which is strictly related to
the second moment of the field distribution. Moreover, the line
shape of the superconducting $P(B)$ is asymmetric, as expected for
a field distribution within a reasonably well arranged VL. This
could be questionable, since
%When $T_{c}$ is crossed there is a clear broadening of the  of the
%in the relaxation of the muons spin polarization, signaling the
%formation of a vortex lattice  in the sample. At $T=10$~K, which
%is above $T_{c}$ for a field of 1200~G, a very slow relaxation is
%resent. On the other hand, at $T=2$~K one can see that about half
%of the signal relaxes quite fast.
%%The relaxation rate of the background is temperature-independent.
recently, it has been shown that the irreversibility line in
CaC$_{6}$ coincides with the $H_{c2}(T)$ line,\cite{Welp} which
suggests that the pinning in this sample is very strong and that a
large number of pinning sites are available. In the
 panel (b) of Fig.~\ref{figure1} the magnetic moment of
 CaC$_6$,  measured at the same field used in TF-$\mu $SR measurements, is reported as a
function of temperature in zero field cooling (ZFC) and field
cooling (FC) conditions. In the FC conditions the magnetic moment
is negligible compared to the ZFC one, indicating a lack of field
expulsion. The same was observed in the $%
\mu $SR experiment: we found almost no change in the average muon
precession frequency when entering the superconducting phase. This
is typically an effect of strong pinning, which confirms previous
results.\cite{Welp} The regularity of the VL could be affected by
the presence of pinning in the sample. The degree of the pinning
induced disorder of the flux line lattice (FLL) depends on many
factors, as the strength and the nature of the pinning and  the
value of the external field. However, the asymmetric line-shape of
the measured field distribution in our samples indicates that,
although a disorder of the FLL is present, it is not large enough
to prevent a reliable analysis of the data.

As a first step,  the $\mu$SR time spectra collected below $T_c$
were fitted by two Gaussian lines :\cite{Khasanov}

%\begin{equation}
\begin{eqnarray}
 \label{twoGaussian}
P(t)=A_{bg}\exp(-\sigma_{bg}^2t^2/2)\cos(\gamma_{\mu}B_{bg}t+\phi)+\\
\nonumber
 A\exp(-\sigma^2 t^2/2)\cos(\gamma_{\mu}Bt+\phi)
%\end{equation}
\end{eqnarray}

\noindent where $A_{bg}, \sigma_{bg}, B_{bg}$ and $A, \sigma, B$
are the asymmetry, the relaxation rate and the mean field of the
background signal and of the superconducting signal, respectively.
The first one is assumed to be temperature- and field-independent.
The second one is an approximation to the field distribution in
the superconducting state of the sample. This distribution is both
temperature- and field-dependent, and it generates a Gaussian
relaxation rate $\sigma(T) $ of the muon spin polarization, which
is proportional to the second moment of the local field
distribution. The minimum field for which we could obtain a
reliable value for $\sigma $ is 60~mT. Below this field the
combination of a very fast relaxation and a relatively low
frequency prevents a good fit.

\begin{figure}[tbp]
%\vspace{0.5cm}
\par
\begin{center}
\includegraphics[width=8cm]{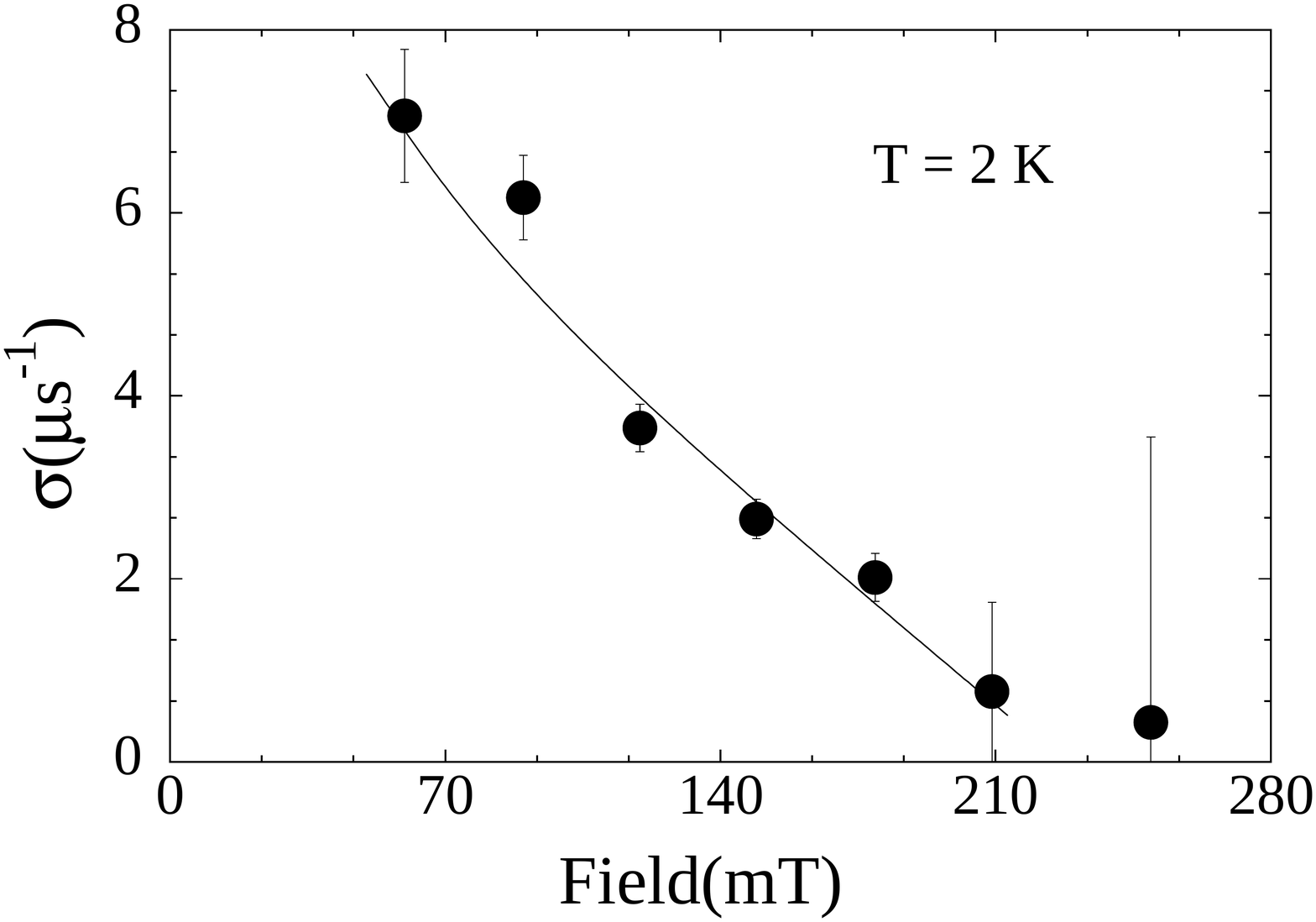}
\end{center}
\caption{Gaussian relaxation as a function of magnetic field at T
= 2 K. The solid line is a fit by Eq. \ref{Brandt}} \label{H_dep}
\end{figure}

 In Fig.~\ref{H_dep} we show the field dependence of $\sigma $
measured at $T\simeq2$~K. The measurements were done in
field-cooled conditions: for every field value the sample was
warmed to a temperature above $T_{c}$ and cooled back to about
2~K. As can be seen, there is an almost linear decrease in the
relaxation rate with increasing field.

For a fully gapped BCS superconductor, as CaC$_6$ is supposed to
be,\cite{Theory,CP, Lamura, Bergeal}  the magnetic penetration
depth $\lambda$ should be field independent.\cite{Landau,Kadono}
In this case, by the analysis of the field dependence of $\sigma$
it is possible to obtain an estimate of the absolute value of
$\lambda$, by using the approximation developed by Brandt [Eq.(13)
in Ref.\onlinecite{Brandt03}], which is considered a very good one
for superconductors with $\kappa = \lambda/\xi \geq$ 5 ($\xi$ is
the coherence length):

\begin{equation}
 \label{Brandt}
 \sigma \approx 0.172\gamma_{\mu} \frac{\Phi_0}{2\pi}  \left(1-\frac{B}{B_{c2}}\right)
 \left[1+1.21\left(1-\sqrt{\frac{B}{B_{c2}}}\right)^3\right]\lambda_{ab}^{-2}
\end{equation}
%\begin{equation}
% \times     \lambda_{ab}^{-2},
%
%\end{equation}

\noindent where $\gamma_{\mu} = 2\pi\times135.5342$~MHz/T is the
muon gyromagnetic ratio, $\Phi_{0}$ is the flux quantum,
%$b$ = B/B$_{c2}^{ab}$
B$_{c2}^{ab}$ the in plane upper critical field, and
$\lambda_{ab}$ the in plane magnetic field penetration depth. The
model reproduces the data reasonably well (solid line in
Fig.~\ref{H_dep}). This is thus consistent with a field
independent $\lambda_{ab}$ and, therefore, with  CaC$_6$ being a
fully gapped superconductor.
 The fit (done without considering the
data point at 250 mT which is too close or above $B_{c2}$ and is
compatible with the background value) gives $\lambda_{ab}$ = 77(3)
nm and $B_{c2}$ = 230(10)mT. The latter, within the
Ginzburg-Landau picture, corresponds to $\xi_{ab} \simeq$ 38 nm.
Both the obtained parameters, $\xi_{ab}$ and $\lambda_{ab}$, are
in  good agreement with previous results
\cite{Emery,CP,Bergeal,ANL_mag,Jobiliong,Lamura,Cubitt} and give
$\kappa = \lambda_{ab}/\xi_{ab} \approx $ 2.  Although the
$\kappa=\lambda/\xi=2$ is rather small the Eq.~\ref{Brandt} is
still valid in the fields $\sqrt{B/B_{c2}}\geq 0.5$ (see Fig.~6 in
Ref.\onlinecite{Brandt03}), that is, in the range of our data.

These results have been obtained by using a gaussian approximation
for the field distribution $P(B)$. To obtain a more accurate value
of $\lambda_{ab}$ in the following we will analyze the muon data
by using the analytical Ginzburg-Landau (AGL) approximation
,\cite{Hao,Yaouanc} which is valid for a broad range of fields and
of Ginzburg-Landau parameter $\kappa$.

The spatial field distribution {\bf B(r)} in the vortex lattice
was calculated as follows:
\begin{equation}
B({\bf{r}}) = \langle B \rangle \sum_{\bf{G}} \exp(-i{\bf{G \cdot
r}}) B_{\bf{G}}(\lambda, \xi, \langle B \rangle).
 \label{eq:Field-distribution}
\end{equation}
The $B_{\bf{G}}$ are the Fourier components:\cite{Hao,Yaouanc}
\begin{equation}\label{eq:BkAGL}
B_{\bf G} = \frac{\Phi_0}{S}\frac{f_\infty
K_1[\frac{\xi_v}{\lambda}(f_\infty^2 + \lambda^2G^2)^{1/2}
]}{(f_\infty^2 +
\lambda^2G^2)^{1/2}K_1(\frac{\xi_v}{\lambda}f_\infty)},
\end{equation}
with $f_\infty = 1-b^4$, and
\[ \xi_v = \xi(\sqrt{2} - \frac{0.75}{\kappa})(1+b^4)^{1/2}[1 - 2b(1-b)^2]^{1/2}. \]
Here, $\langle B \rangle$ is the average magnetic field inside the
superconductor, $\xi$ the coherence length,  $K_1(x)$ is the
modified Bessel function, $\bf{r}$ the vector coordinate in a
plane perpendicular to the applied field, ${\bf{G}} =
4\pi/\sqrt{3}a(m\sqrt{3}/2, n+m/2)$ the reciprocal lattice vector
of the hexagonal FLL lattice, $a$ the intervortex distance, and
$m$ and $n$ are integer numbers. From this spatial field
distribution the probability field distribution $P(B)$ was
calculated.

The $\mu$SR time spectra were fitted with the depolarization
function P(t) obtained from P(B) as follows:

\begin{align}\label{eq:TdomainSpec}
P(t) =& A\cdot e^{-1/2\sigma_g^2t^2+i\phi}\int P(B)e^{i\gamma_\mu B t}dB +\\
& A_{bg}\cdot e^{-\sigma_L t+i(\gamma_\mu B_{a}t + \phi)}  \,
,\nonumber
\end{align}
where the second term is the signal of the background with
asymmetry $A_{BG}$ and relaxation $\sigma_L$. The parameter
$\sigma_g$ describes the disorder of the FLL and was fixed
proportional to $1/\lambda^2$ such that
$\sigma_{g}(0)=2.1$~$\mu$s$^{-1}$ (note that this relaxation is
 much smaller than the superconducting relaxation, which are
summed in quadrature). Such relation corresponds to the rigid
(well pinned) FLL.\cite{Riseman95, Mais} We neglected additional
nuclear relaxation, since the concentration of Ca and C isotopes
with magnetic nuclear moments is practically zero. The parameter
$\xi$ was kept fixed during the fit at the values determined, at
each temperature, from the corresponding $B_{c2}$(T)  via the
Ginzburg-Landau relation $\xi(T) = \sqrt(\Phi_0/(2\pi
B_{c2}(T)))$. The $B_{c2}$(T) values were obtained by linearly
interpolating the following four experimental points:
$B_{c2}(2K)=230$~mT, given above by the fit of the field
dependence of $\sigma$, and $B_{c2}(5.6K)=120$~mT,
$B_{c2}(8.5K)=60$~mT, $B_{c2}(11.5K)=0$ measured from
magnetization experiments (only the measurement at 120 mT is shown
in Fig.~\ref{figure1}). These experimental points agree well with
the $B_{c2}(T)$ curve reported by Cubitt \textit{et
al.}.\cite{Cubitt} This model represents very accurately the
experimental data, as shown by the solid line in Fig.~\ref{FFT}.

\begin{figure}[tbp]
%\vspace{0.5cm}
\par
\begin{center}
\includegraphics[width=8cm]{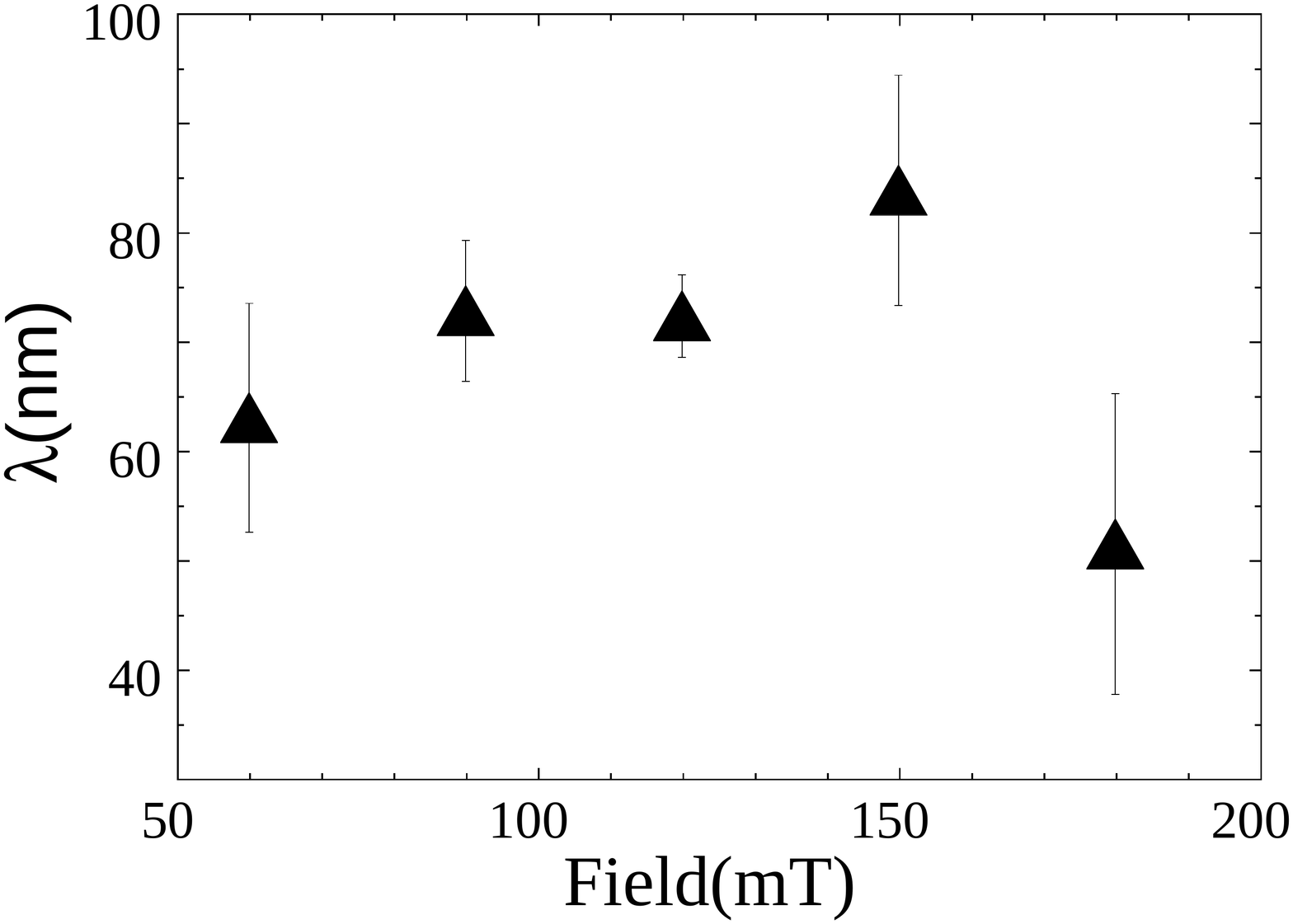}
\end{center}
\caption{Magnetic penetration depth as a function of magnetic
field obtained from the fit to the $\mu$SR time spectra at  2 K.}
\label{lambdaVSb}
\end{figure}

In Fig.~\ref{lambdaVSb}, the values of $\lambda_{ab}$, obtained by
fitting the $\mu$SR data with Eq.~\ref{eq:TdomainSpec} at the
lowest temperature and several fields, are shown.
No field dependence of $\lambda_{ab}$ is detected within
error-bar, as expected for conventional fully gapped BCS
superconductors and already suggested by the second moment
analysis (Fig.~\ref{H_dep}). A weighted average of the data gives
$\lambda_{ab}$ = 72(3) nm, compatible, within the errors, with the
value obtained by fitting the field dependence of the relaxation
rate, and in very good agreement with the value reported in Ref.
\onlinecite{Lamura}. Recently Cubitt {\it et al.} \cite{Cubitt}
reported a magnetic penetration depth $\lambda = 50$~nm using SANS
on CaC$_6$ compound. In their analysis they used the London model
with Gaussian  cut off (LGC). As it was shown previously by
Yaouanc et al.,\cite{Yaouanc} the form factors of the LGC model
deviate substantially from  the exact solution of the
Ginzburg-Landau model in a very broad range of fields. Indeed, an
analysis of our data by using the LGC model gives $\lambda_{ab} =
52(4)$~nm, in good agreement with the results of Cubitt {\it et
al.} \cite{Cubitt} The AGL model used here gives more reliable
results ,\cite{Yaouanc} and systematic deviations of the AGL model
from the exact solution of the Ginzburg-Landau \cite{Brandt97}
model are within the error bar.

The magnetic penetration depth is defined, in the London limit of
low fields, as a measure of the superfluid density $\rho \propto
\lambda^{-2}$. However, for fields close to $B_{c2}$ the mean
magnitude of the Ginzburg-Landau order parameter $\psi(r)$ drops
due to a substantial overlapping of the vortex
cores.\cite{Brandt97} The spatial average of the superfluid
density is thus well described by:\cite{Maisuradze09d, Lipavski}

\begin{equation}
\rho(T) \propto \langle\mid\psi(r)\mid^2 \rangle \lambda^{-2}(T)
\simeq (1-B/B_{c2}(T))\lambda^{-2}(T),
 \label{superfldens}
\end{equation}%

\noindent where $\langle \cdot \cdot \cdot \rangle$ refers to the
spatial average over the unit cell of the FLL. The temperature
dependence of $\lambda$, needed to calculate $\rho(T)$ with
Eq.\ref{superfldens}, was obtained by fitting of
Eq.~\ref{eq:TdomainSpec} to the data collected at different
temperatures in a field of 120 mT. The $B_{c2}(T)$ values were
obtained as described above.

\begin{figure}[tbp]
%\vspace{0.5cm}
\par
\begin{center}
\includegraphics[width=8cm]{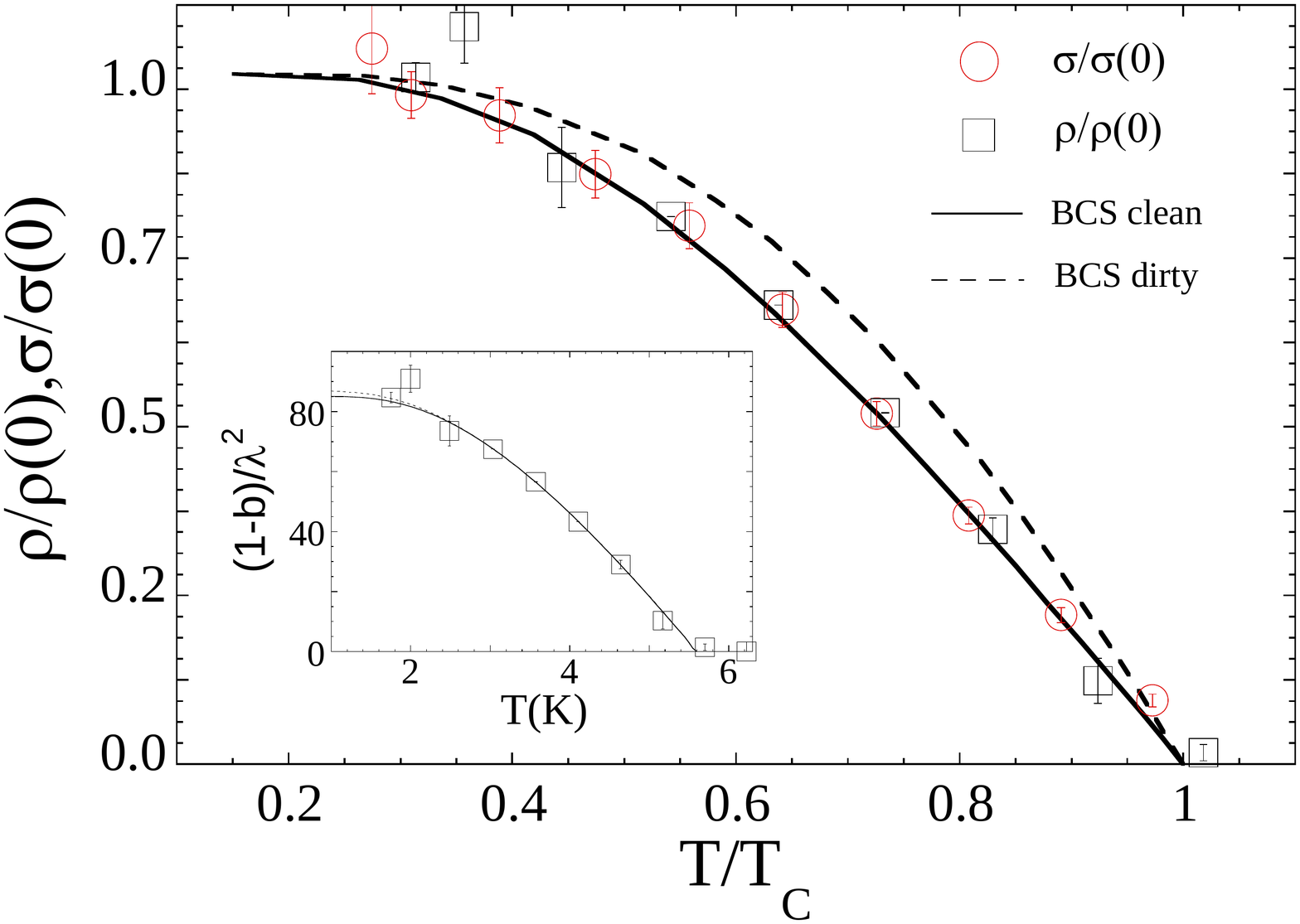}
\end{center}
\caption{Normalized superfluid density $\rho(T)/\rho(0)$ and muon
spin relaxation rate $\sigma(T)/\sigma(0)$ as a function of the
reduced temperature $T/T_c$ measured at 120 mT. Full and dashed
lines are BCS clean limit and dirty limit curves, respectively.
Inset: temperature dependence of $\rho$ and BCS fits.}
\label{rhoVSt}
\end{figure}
In Fig.~\ref{rhoVSt}, we show the normalized superfluid density
$\rho(T)/\rho(0)$ as a function of the reduced temperature $T/T_c$
measured at 120 mT. $\rho(0)$ is obtained from the BCS fit to the
temperature dependent $\rho(T)$ by using the
equation:\cite{tinkham}

\begin{equation}
\rho(T)=\rho(0)[1- \frac{2}{k_{B}T}
\int_{0}^{\infty}f(\varepsilon,T,\Delta(T))[1-f(\varepsilon,T,\Delta(T))]d\varepsilon]
\label{BCS}
\end{equation}
where $f(\varepsilon,T,\Delta(T))$ =
$[1+\exp[\sqrt{\varepsilon^2+\Delta^2(T)}/(k_B T_C)]]^{-1}$ is the
Fermi distribution and $\Delta(T)=\Delta(0)\delta(T/T_c)$, with
$\delta(T/T_c)$ given by the conventional BCS  temperature
dependence of the gap.\cite{Muhlschlegel} Indeed, although data at
temperatures lower than 1.75 K would be important to clearly
assess the occurrence of a saturation, however the absence of a
field dependence of $\lambda$ points to nodeless gap in the
$ab$-plane of CaC$_6$, which is consistent with previous
theoretical end experimental studies.\cite{Theory,CP, Lamura,
Bergeal} Therefore it is correct to fit the experimental data
using Eq. \ref{BCS}. In order to reduce the number of fitting
parameters, the fit has been performed by keeping $T_c$ fixed at
the value 5.6(1) K obtained from magnetization data at the same
field of 120 mT (see Fig.~\ref{figure1}). As shown in the inset to
Fig.~\ref{rhoVSt}, the fitting curve (solid line) represents well
the experimental data, yielding $\rho(0)$ = 85.1(3) $\mu$m$^{-2}$
and $\Delta(0)$ = 0.868(5) meV, which gives 2$\Delta(0)$/k$_BT_c$
= 3.6(1). This value is compatible with the one expected for a
weak coupling BCS superconductor and is in good agreement with
previous results.\cite{CP, Lamura, Bergeal} A fit with the dirty
limit BCS temperature dependence \cite{tinkham} is also acceptable
(see inset to Fig.~\ref{rhoVSt}), although with larger $\chi ^2$
and larger errors of the fitting parameters ($\rho(0)$ = 87(2)
$\mu$m$^{-2}$, $\Delta(0)$ = 0.66(2) meV). With these parameters
we would obtain 2$\Delta$/k$_BT_c$ = 2.74(1). This quantity is
substantially different from the previously reported ones
\cite{CP, Lamura, Bergeal} and it is well below the BCS
weak-coupling ratio. The normalized superfluid density, shown in
the main panel of Fig.~\ref{rhoVSt}, clearly follows the clean
limit curve, and is not compatible with the dirty limit one,
suggesting that bulk CaC$_6$ is a BCS superconductor in the clean
limit regime, in agreement with Mialitsin \textit{et
al.}\cite{Mialitsin} and Cubitt \textit{et al.}\cite{Cubitt}

In the main panel of Fig.~\ref{rhoVSt}  the temperature dependence
of the normalized relaxation rate $\sigma$(T), obtained by the two
gaussian model (Eq.~\ref{twoGaussian}), is also reported. These
data closely resemble the $\rho(T)$ curve. Indeed, as follows from
the Fig.~6 of Ref.~\onlinecite{Brandt03} the quantity $\sigma
\kappa^2 /B_{c2}(1-b) \propto \sigma/[(1-b)/\lambda^2]$ is
practically independent of the reduced field $b = \langle
B\rangle/B_{c2}$ for $\kappa \simeq 2$ and $\sqrt{b}>0.25$. On the
other hand, $\sigma/[(1-b)/\lambda^2]$ is proportional to $\sigma
/ \rho$. Therefore, the present case with $\kappa \simeq 2$ is
exceptional, when a simple second moment analysis gives a quite
precise result for the temperature dependence of the superfluid
density in a broad range of fields.

To summarize, we performed TF-$\mu $SR measurements on the
intercalated graphite superconductor CaC$_{6}$ in the
vortex-lattice state. Our experimental method included the use of
Kapton tape to gather good data from very thin samples, ~50$~\mu
$m thick.  Despite of strong pinning, we detected an asymmetric
field distribution, typically observed in superconducting single
crystals in the vortex state. The data were analyzed with two
different models. Using a simple two-Gaussian model, the
relaxation rate $\sigma$ was obtained. Its field dependence
suggested a field independent $\lambda_{ab}(0)$ = 77(3) nm and a
coherence length $\xi_{ab}(0) \simeq$ 38 nm. The more precise
analytical Ginzburg-Landau model yielded $\lambda_{ab}(0)$ = 72(3)
nm, a value remarkably close to that one obtained by the second
moment analysis and  in good agreement with previous microwave
measurements.\cite{Lamura} Also this analysis reveal that
$\lambda_{ab}(0)$ does not depend on magnetic field, suggesting a
fully gapped superconductivity in CaC$_6$, in agreement with
previous reports.\cite{CP, Lamura, Bergeal} The temperature
dependence of the superfluid density was analyzed by both the
clean and the dirty limit BCS models. The experimental data are
well fitted by the clean limit one, giving a gap-to-$T_c$ ratio
2$\Delta(0)$/k$_BT_c$ = 3.6(1), in agreement with previous results
obtained by surface sensitive techniques.\cite{Lamura, Bergeal}
The temperature dependence of the normalized superfluid density
results to be identical to the temperature dependence of the
normalized relaxation rate, as it should be for a $\kappa \simeq
2$ superconductor at fields $B>0.1\cdot B_{c2}$. Both are well
represented by the clean limit BCS curve,  suggesting that CaC$_6$
is an s-wave  BCS superconductor in the clean limit regime.

This work was partly performed at the Swiss Muon Source (S$\mu$S), Paul Scherrer Institute (PSI, Switzerland).
The authors are grateful to R. Khasanov for useful discussions. This work was partly supported by the EU Project
CoMePhS, by the Israel Science Foundation and by the European Commission under the 6$^{th}$ Framework Programme
through the Key Action: Strengthening the European Research Area, Research Infrastructures; contract no:
RII3-CT-2003-505925. The work at  Argonne National Laboratory was supported by UChicago Argonne, LLC, Operator
of Argonne National Laboratory ("Argonne"). Argonne, a U.S. Department of Energy Office of Science laboratory,
is operated under Contract No. DE-AC02-06CH11357.


\begin{thebibliography}{99}
\bibitem{note} The measurements were performed into two runs. The second run was
needed in order to have additional points at 1.75 K and 5.7 K  in
the temperature dependence of the superfluid density. Also the
point at 2 K was measured a second time, and the point shown in
this manuscript comes from the average of the parameters obtained
by the fit of the two runs.
\end{thebibliography}
\end{document}